\documentclass[12pt,cite]{article}

\newcommand{\Journal}[2]{#1 #2}
\newcommand{\Author}[2]{#1.#2}
\newcommand{\Title}[5]{{\it#1}#2 {\bf#3} (#5) #4}

\newcommand{\PR}{Phys.Rev.}
\newcommand{\NP}{Nucl.Phys.}
\newcommand{\PL}{Phys.Lett.}
\newcommand{\JETP}{JETP}
\newcommand{\SPAWPM}{Sitzungsb. Preu{\ss}. Akad. Wiss. Phys. Math.}
\newcommand{\IJMP}{Int.J.Mod.Phys.}

\begin{document}

\centerline{{\Large {\bf Neutrino Spin Evolution in Presence}}}
\centerline{{\Large {\bf of General External Fields}}}

\centerline{{\large {\bf M.Dvornikov
\footnote{e-mail: maxim\_dvornikov@aport.ru}
A.Studenikin
\footnote{e-mail: studenik@srd.sinp.msu.ru}}}}

\centerline{{\bf Department of Theoretical Physics,
Moscow State University,}}
\centerline{{\bf 119992 Moscow, Russia}}

\centerline{{\it Abstract}}
{\it
The derivation of the quasiclassical Lorentz invariant neutrino spin
evolution equation taking into account general types of neutrino
non-derivative interactions with external fields is presented.
We discuss the
constraints on the characteristics of matter and neutrino
under which this quasiclassical approach is valid.
The application of the obtained equation for the case of the Standard
Model neutrino interactions with moving and polarized
background matter is considered.
}

It is commonly believed that neutrino physics provides strong
evidence for physics beyond the Standard Model. In different
extensions of the Standard Model new types of interactions are
predicted for massive neutrinos.
The problem of neutrino propagation in matter in the case of
a general set of
interactions with the background fermions has attracted
considerable attention (see, for example
\cite{Nu.Se.Sm.Va/NP(97),Be.Gr.Na(99)}). Recently, we have developed
the Lorentz invariant formalism  for description of neutrino
spin-flavor
and flavor oscillations with the Standard Model vector and
axial-vector interactions in moving matter under the influence of
an arbitrary electromagnetic fields
\cite{Eg.Lo.St/HEP-PH(9902447),Lo.St/PL(01),Gr.Lo.St/HEP-PH(0112304)}.
In particular, we have derived, within general assumption like
Lorentz invariance and linearity over neutrino spin vector $S^{\mu}$
and also over such characteristics of matter like fermions currents
and polarizations, the evolution equation for the neutrino spin.
We have used
this evolution equation for description of neutrino
oscillations in electromagnetic fields accounting for neutrino vector
and axial-vector interactions with background fermions that
corresponds to the case of the Standard Model weak interactions.
Note that quasiclassical approach to the problem of neutrino spin
relaxation in stochastic electromagnetic fields was used in
ref.\cite{Lo.St/PR(89)}.
However, the problem of the neutrino spin evolution equation
accounting also for more general
new types of neutrino interactions is still
remained open.

We discuss below neutrino spin evolution in background matter in the
case of a new physics model that admits a general set of new neutrino
interactions.
The goal of this paper is to derive the neutrino spin evolution
equation starting directly from the neutrino
interaction Lagrangian. We suppose that the neutrino
interaction Lagrangian includes not only the Standard Model
vector and axial-vector terms, but also non-derivative
scalar, pseudoscalar,
tensor and pseudotensor interactions.

The derivation of the neutrino spin evolution equation presented
here is based on general spin evolution equation in the Heisenberg
representation.
This approach allows us to analyse more attentively
contributions to the neutrino spin evolution of different mentioned
above external fields.

To derive the neutrino
quasiclassical spin evolution equation we start with the
quantum equation in the Heisenberg representation which describes
spin evolution of a fermion having an energy
$E_{\nu}$, momentum $\vec p$ and mass $m_{\nu}$
(see, for instance, \cite{Te/JETP(90)})
$${\dot {\vec O}}=i[{\cal H},{\vec O}]_{-}. \eqno(1)$$
The spin operator is determined as
$${\vec O}=\rho_{3}{\vec \Sigma}+\rho_{1}{{\vec p}\over{E_{\nu}}}-
\rho_{3}
{{{\vec p}({\vec p}{\vec \Sigma})}\over{E_{\nu}(E_{\nu}+m_{\nu})}},
\eqno(2)$$
where $\rho_{1}=-\gamma^{5}$, $\rho_{3}=\gamma^{0}$,
${\vec \Sigma}=\gamma^{0}\gamma^{5}{\vec \gamma}$.
Note that here we take $\hbar=c=1$.
The Hamiltonian ${\cal H}$ in eq.(1) describes time behavior of
four-component neutrino wave function ${\nu}(x)$.

The Lagrangian $\cal L$ accounting for general types
of neutrino non-derivative interactions with external
fields is chosen in following form,
$$
\matrix{
-{\cal L}=g_{s}s(x){\bar \nu}\nu+
g_{p}{\pi}(x){\bar \nu}\gamma^{5}\nu+
g_{v}V^{\mu}(x){\bar \nu}\gamma_{\mu}\nu+
g_{a}A^{\mu}(x){\bar \nu}\gamma_{\mu}\gamma^{5}\nu+
\cr
+{{g_{t}}\over{2}}T^{\mu\nu}{\bar \nu}\sigma_{\mu\nu}\nu+
{{g^{\prime}_{t}}\over{2}}
\Pi^{\mu\nu}{\bar \nu}\sigma_{\mu\nu}\gamma_{5}\nu,
\cr}
\eqno(3)$$
where $s, \pi, V^{\mu}=(V^{0}, {\vec V}), A^{\mu}=(A^{0}, {\vec A}),
T_{\mu\nu}=({\vec a}, {\vec b}), \Pi_{\mu\nu}=({\vec c}, {\vec d})$
are the scalar, pseudoscalar, vector, axial-vector, tensor,
pseudotensor fields, respectively, and
$g_{i}$ $(i=a, p, v, a, t, t^{\prime})$
are corresponding coupling constants,
$\sigma_{\mu\nu}={{i}\over{2}}(\gamma_{\mu}\gamma_{\nu}-
\gamma_{\nu}\gamma_{\mu})$.
With the use of the Lagrangian of eq.(3)
the expression for the Hamiltonian $\cal H$ is straightforward:
$$
\matrix{
{\cal H}=g_{s}s\rho_{3}-
ig_{p}\pi\rho_{2}+
g_{v}(V^{0}-({\vec \alpha}{\vec V}))-
g_{a}(\rho_{1}A^{0}-({\vec \Sigma}{\vec A}))-
\cr
-g_{t}(\rho_{3}({\vec \Sigma}{\vec b})+
\rho_{2}({\vec \Sigma}{\vec a}))-
ig^{\prime}_{t}(\rho_{3}({\vec \Sigma}{\vec c})-
\rho_{2}({\vec \Sigma}{\vec d})),
\cr} \eqno(4)$$
where ${\vec \alpha}=\gamma^{0}{\vec \gamma}$,
$\rho_{2}=i\rho_{1}\rho_{3}$. Then, from eq.(1) we obtain:
$$
\matrix{
{\dot {\vec O}}=
-2g_{s}s{{\vec p}\over{E_{\nu}}}+
2ig_{p}\pi
\left\{
\rho_{1}{\vec \Sigma}-\rho_{3}{{\vec p}\over{E_{\nu}}}-
\rho_{1}
{{{\vec p}({\vec p}{\vec \Sigma})}\over{E_{\nu}(E_{\nu}+m_{\nu})}}
\right\}-
\cr
-2g_{v}\rho_{2}
\left\{
{\vec V}-
{{{\vec p}({\vec p}{\vec V})}\over{E_{\nu}(E_{\nu}+m_{\nu})}}
\right\}-
\cr
-2g_{a}
\left\{
A^{0}\rho_{2}
\left(
{\vec \Sigma}-
{{{\vec p}({\vec p}{\vec \Sigma})}\over{E_{\nu}(E_{\nu}+m_{\nu})}}
\right)-
\rho_{3}
\left(
{\vec A}\times{\vec \Sigma}+
{{{\vec p}({\vec A}[{\vec p}\times{\vec \Sigma}])}
\over{E_{\nu}(E_{\nu}+m_{\nu})}}
\right)
\right\}+
\cr
+2g_{t}\left\{
[{\vec \Sigma}\times{\vec b}]+
\rho_{2}{{\vec p}\over{E_{\nu}}}({\vec \Sigma}{\vec b})+
{{{\vec p}({\vec \Sigma}[{\vec p}\times{\vec b}])}
\over{E_{\nu}(E_{\nu}+m_{\nu})}}+
\rho_{1}
\left(
{\vec a}-
{{{\vec p}({\vec p}{\vec a})}\over{E_{\nu}(E_{\nu}+m_{\nu})}}
\right)-
\rho_{3}{{\vec p}\over{E_{\nu}}}({\vec \Sigma}{\vec a})
\right\} +
\cr
+2ig^{\prime}_{t}\left\{
[{\vec \Sigma}\times{\vec c}]+
\rho_{2}{{\vec p}\over{E_{\nu}}}({\vec \Sigma}{\vec c})+
{{{\vec p}({\vec \Sigma}[{\vec p}\times{\vec c}])}
\over{E_{\nu}(E_{\nu}+m_{\nu})}}-
\rho_{1}
\left(
{\vec d}-
{{{\vec p}({\vec p}{\vec d})}\over{E_{\nu}(E_{\nu}+m_{\nu})}}
\right)+
\rho_{3}{{\vec p}\over{E_{\nu}}}({\vec \Sigma}{\vec d})
\right\}.
\cr}
\eqno(5)$$
In getting this equation we suppose that
all external fields are independent of spatial coordinates.

It should be noted that
the equation obtained does not seem to have classical
interpretation because of the {\it zitterbewegung}
\cite{Sh/SPAWPM(30)}.
To eliminate
this phenomenon we extract, following
the recipe of the ref.\cite{Te/JETP(90)}, an even part from eq.(5):

$$\{{\dot {\vec O}}\}=
{{1}\over{2E_{\nu}}}
\{{\dot {\vec O}},{\cal H}_{0}\}_{+}, \eqno(6)$$
where ${\cal H}_{0}={\vec \alpha}{\vec p}+\rho_{3}m_{\nu}$.

Performing anticommutations in eq.(6) we get the following
expression:
$$
\matrix{
\{{\dot {\vec O}}\}=
2g_{a}
\left(
{{A^{0}}\over{E_{\nu}}}[{\vec O}\times{\vec p}]-
{{m_{\nu}}\over{E_{\nu}}}[{\vec O}\times{\vec A}]-
{{1}\over{E_{\nu}(E_{\nu}+m_{\nu})}}
({\vec A}{\vec p})[{\vec O}\times{\vec p}]
\right)+
\cr
+2g_{t}\left(
[{\vec O}\times{\vec b}]-
{{1}\over{E_{\nu}(E_{\nu}+m_{\nu})}}
({\vec p}{\vec b})[{\vec O}\times{\vec p}]+
{{1}\over{E_{\nu}}}[{\vec O}\times[{\vec a}\times{\vec p}]]
\right)+
\cr
+2ig^{\prime}_{t}\left(
[{\vec O}\times{\vec c}]-
{{1}\over{E_{\nu}(E_{\nu}+m_{\nu})}}
({\vec p}{\vec c})[{\vec O}\times{\vec p}]-
{{1}\over{E_{\nu}}}[{\vec O}\times[{\vec d}\times{\vec p}]]
\right).
\cr}
\eqno(7)$$
In the quasiclassical approximation ($\hbar\to 0$) for any operator
$\hat F$ one has ${\dot {\hat F}}=\{{\dot {\hat F}}\}$
\cite{Te/JETP(90)}. That is why
in order to obtain the quasiclassical equation for the neutrino
spin
evolution we substitute $\{{\dot {\vec O}}\}$ for ${\dot {\vec O}}$.
Then, averaging the relevant equation over the stationary neutrino
wave function and taking into account that
$$<{\vec O}>={\vec \zeta}_{\nu},
\quad <{\vec p}>={\vec \beta}E_{\nu},$$
where ${\vec \beta}$ is the neutrino velocity,
we obtain the relativistic equation for the neutrino spin
vector ${\vec \zeta}_{\nu}$ in the form
$$
\matrix{
{\dot {\vec \zeta}_{\nu}}=
2g_{a}\left\{
A^{0}[{\vec \zeta}_{\nu}\times{\vec \beta}]-
{{m_{\nu}}\over{E_{\nu}}}[{\vec \zeta}_{\nu}\times{\vec A}]-
{{E_{\nu}}\over{E_{\nu}+m_{\nu}}}
({\vec A}{\vec \beta})[{\vec \zeta}_{\nu}\times{\vec \beta}]
\right\}+
\cr
+2g_{t}\left\{
[{\vec \zeta}_{\nu}\times{\vec b}]-
{{E_{\nu}}\over{E_{\nu}+m_{\nu}}}
({\vec \beta}{\vec b})[{\vec \zeta}_{\nu}\times{\vec \beta}]+
[{\vec \zeta}_{\nu}\times[{\vec a}\times{\vec \beta}]]
\right\}+
\cr
+2ig^{\prime}_{t}\left\{
[{\vec \zeta}_{\nu}\times{\vec c}]-
{{E_{\nu}}\over{E_{\nu}+m_{\nu}}}
({\vec \beta}{\vec c})[{\vec \zeta}_{\nu}\times{\vec \beta}]-
[{\vec \zeta}_{\nu}\times[{\vec d}\times{\vec \beta}]]
\right\}.
\cr}
\eqno(8)$$
It is worth to be noted that in agreement with
\cite{Be.Gr.Na(99)} neither scalar nor pseudoscalar nor vector
interaction contributes to the neutrino spin evolution.

The Lorenz invariant form of eq.(8) can be obtained
using the four-dimensional spin vector $S^{\mu}$ which is determined
by the three-dimensional spin vector
${\vec \zeta}_{\nu}$
in accordance with the relation:
$$S^{\mu}=
\left(
{{({\vec \zeta}_{\nu}<{\vec p}>)}\over{m_{\nu}}},
{\vec \zeta}_{\nu}+
{{<{\vec p}>({\vec \zeta}_{\nu}<{\vec p}>)}
\over{m_{\nu}(m_{\nu}+E_{\nu})}}
\right).
\eqno(9)$$
Thus, we get the Lorentz invariant form for the neutrino spin
$S^{\mu}$ evolution
equation accounting for the general interactions with external
fields
$${{dS^{\mu}}\over{d\tau}}=2g_{t}
(T^{\mu\nu}S_{\nu}-
u^{\mu}T^{\lambda\rho}u_{\lambda}S_{\rho})+
2ig^{\prime}_{t}
({\tilde \Pi}^{\mu\nu}S_{\nu}-
u^{\mu}{\tilde \Pi}^{\lambda\rho}u_{\lambda}S_{\rho})
+2g_{a}G^{\mu\nu}S_{\nu},
\eqno(10)$$
where
$G^{\mu\nu}=\varepsilon^{\mu\nu\alpha\beta}A_{\alpha}u_{\beta}$,
$u^{\mu}={{E_{\nu}}\over{m_{\nu}}}(1,{\vec \beta})$,
${\tilde \Pi}^{\mu\nu}=
{{1}\over{2}}\varepsilon^{\mu\nu\alpha\beta}\Pi_{\alpha\beta}$.
The tensor $G_{\mu\nu}$ can be expressed through two vectors
$G_{\mu\nu}=(-{\vec P},{\vec M})$ which are presented in the form,
$$
\matrix{
{\vec M}=\gamma(A^{0}{\vec \beta}-{\vec A}),
\cr
{\vec P}=-\gamma[{\vec \beta}\times{\vec A}],
\cr} \eqno(11)$$
where $\gamma={{E_{\nu}}\over{m_{\nu}}}.$
The derivation in the left-handed side of eq.(10) is taken over the
neutrino proper time $\tau ={{m_{\nu}}\over{E_{\nu}}}t$, where $t$
is the time in the laboratory frame of reference.

The neutrino spin evolution equation (10) can
be used for any theoretical model in which neutrino has mentioned
above general interactions.
Let us consider now the case of the Standard Model neutrino
interactions which are sure to be one of the possible applications of
the approach concerned and suppose that matter is composed of
electrons, neutrons and protons. Then, the coupling
constants entering in the Lagrangian of eq.(3) are $g_{i}=0$ (for
$i=s, p, t^{\prime}$), $g_{v}=g_{a}={{G_{F}}\over{\sqrt{2}}}$,
$g_{t}=\mu$, where
$G_{F}$ is the Fermi constant and $\mu$ is the neutrino
magnetic moment.  The vector and axial-vector fields are expressed
in the form (see, for instance \cite{Pa/IJMP(92)})
$$A^{\mu}=V^{\mu}=\sum\limits_{f=e,p,n}
j^{\mu}_{f}q^{(1)}_{f}+\lambda^{\mu}_{f}q^{(2)}_{f}
\eqno(12)$$
where
$$
\matrix{
q^{(1)}_{f}=
(I_{3L}^{(f)}-2Q^{(f)}\sin^{2}\theta_{W}+\delta_{ef}),
\quad
q^{(2)}_{f}=-(I_{3L}^{(f)}+\delta_{ef}),
\cr
\delta_{ef}=\left\{
\begin{tabular}{l l}
1 & {\it f=e} \\
0 & {\it f=n, p} \\
\end{tabular}
\right.
\cr} \eqno(13)$$
and $I_{3L}^{(f)}$ is the
value of the isospin third component of a fermion $f$,
$Q^{(f)}$ is the value of its electric charge and $\theta_{W}$ is
the Weinberg angle.
In the case of the Standard Model the tensor field corresponds to
the electromagnetic interaction and
$$T_{\mu\nu}=F_{\mu\nu}=({\vec E},{\vec B}), \eqno(14)$$
where $\vec E$ and $\vec B$ are the electric and magnetic fields
respectively.
The summation in eq.(12) is performed over the background
electrons, protons and neutrons. Eq.(12)
for external field $A^{\mu}$ depends upon fermions currents
$j^{\mu}_{f}$ and polarizations $\lambda^{\mu}_{f}$ which are related
with the matter components number
densities $n_{f}$, speeds ${\vec v}_{f}$ of the reference frames
in which the mean momentum of each of the
fermions $f=e, p, n$
equals zero and the mean value
of the fermions
polarization vectors ${\vec \zeta}_{f}$:
$$
\matrix{
j^{\mu}_{f}=(n_{f},n_{f}{\vec v}_{f})
\cr
{\lambda}^{\mu}_{f}=
\left(n_{f}({\vec \zeta}_{f}{\vec v}_{f}),
n_{f}{\vec \zeta}_{f}\sqrt{1-v^{2}_{f}}
+{{n_{f}{\vec v}_{f}({\vec \zeta}_{f}{\vec v}_{f})}
\over{1+\sqrt{1-v^{2}_{f}}}}
\right).
\cr
} \eqno(15)$$
The details of the averaging procedure over the
background fermions are discussed in
ref. \cite{Lo.St/PL(01)}.

The covariant neutrino spin evolution equation with the external
fields defined by eqs.(10 - 15) enables one to describe the neutrino
spin precession in the case of the Standard Model in the
arbitrary moving and polarized matter, with the neutrino mass
accounted exactly.\footnote{Our approach also does not imply
the use of the assumption
$E_{\nu}\gg m_{\nu}$.}
On the base of the eq.(10) (see also (8))
let us now consider the equation for the
neutrino spin vector ${\vec \zeta}_{\nu}$,
$${\dot {\vec \zeta}_{\nu}}=
{{2\mu}\over{\gamma}}
[{\vec \zeta}_{\nu}\times{\vec B}_{0}]+
{{\sqrt{2}G_{F}}\over{\gamma}}
[{\vec \zeta}_{\nu}\times{\vec M}_{0}], \eqno(16)$$
where
$$
\matrix{
{\vec M}_{0}=
\gamma{\vec \beta}
\left(A^{0}-
{{E_{\nu}}\over{E_{\nu}+m_{\nu}}}
({\vec A}{\vec \beta})\right)-{\vec A},
\cr
{\vec B}_{0}=\gamma\left\{{\vec B}+[{\vec E}\times{\vec \beta}]-
{{E_{\nu}}\over{E_{\nu}+m_{\nu}}}
{\vec \beta}({\vec \beta}{\vec B})\right\},
\cr} \eqno(17)$$
are
the quantities  determined in the neutrino rest frame. The derived
eqs.(16-17) reproduce the spin evolution equations for neutrinos
propagating in the background electrons (the electron plasma) which
were received in ref. \cite{Lo.St/PL(01)}
on the base of the generalization of the
Bargmann-Michel-Telegdi equation. We underline that the approach
considered here implies the use as a starting point the {\it Lorentz
invariant} neutrino interaction Lagrangian for derivation of the {\it
Lorentz invariant} spin evolution equation.

It is worth mentioning that neutrino spin evolution problem in
presence of the background fermions can be considered using kinetic
equations approach \cite{Se/PRD(93)}.  The analog of the eqs.(16-17)
for the specific cases of nonmoving $({\vec v}_{f}=0)$ and
unpolarized $({\vec \zeta}_{f}=0)$ medium was shown to be
derived from the more general kinetic equation for the very important
case of the ultrarelativistic neutrino. However our approach, as it
was mentioned above, is valid in cases of arbitrary speed of
neutrino and arbitrary moving and polarized background matter.

Let us discuss in some details approximations which we use in
deriving the neutrino spin evolution equation.
First we have neglected
spatial coordinate dependence of all external
fields. To analyze the adequacy of this approximation we now consider
the opposite case.
For simplicity, we shall discuss the Standard Model neutrino
interactions in the case of
nonmoving and unpolarized background matter. Then,
the Hamiltonian (4) takes the form,
$${\cal H}=
{{G_{F}}\over{\sqrt{2}}}n_{eff}(1+{\gamma}_{5}), \eqno(18)$$
where $n_{eff}=\sum\limits_{f=e,p,n}n_{f}q_{f}^{(1)}$ is
an effective density of the background matter which is now supposed
to depend on spatial coordinates.
To study corrections to eq.(8) induced by the spatial
dependence of $n_{eff}$
we must take into account the
commutation relation,
$[n_{eff}, {\vec p}]_{-}=i\hbar{\vec \nabla}n_{eff}$. Contrary to the
used above agrement here we set $\hbar\not= 1$.
Thus, holding the first order corrections in expansion over
$\hbar$ we obtain an analog of
eq.(8),
$$
\matrix{
{\dot {\vec \zeta}_{\nu}}=
{{\sqrt{2}G_{F}}\over{\hbar}}
\Big\{
n_{eff}[{\vec \zeta}_{\nu}\times{\vec \beta}]-
\cr
-{{\hbar}\over{2(E_{\nu}+m_{\nu})}}
\bigg[
[[{\vec \beta}\times{\vec \nabla}n_{eff}]\times{\vec \zeta}_{\nu}]+
({\vec \nabla}n_{eff}{\vec \beta})
\bigg(
{{E_{\nu}}\over{E_{\nu}+m_{\nu}}}
({\vec \beta}{\vec \zeta}_{\nu})-1
\bigg)
\bigg]-
\cr
-{{i\hbar}\over{2E}}
[{\vec \nabla}n_{eff}\times<\rho_{3}{\vec \Sigma}>]
\Big\},
\cr} \eqno(19)$$
where the average in the last form
$<\rho_{3}{\vec \Sigma}>=
\int d^{3}{\bf x}\nu^{+}\rho_{3}{\vec \Sigma}\nu$.
It follows that for
the relativistic neutrinos $(E_{\nu}\gg m_{\nu})$
the additional quantum terms
($\sim\hbar$) vanish if the following constrained is fulfilled
$${{\hbar}\over{E_{\nu}}}
\left|{{{\vec \nabla}n_{eff}}\over{n_{eff}}}\right|\ll 1.
\eqno(20)$$
This condition implies very slow effective density variation along
the neutrino wave package width $L={{\hbar}\over{E_{\nu}}}$.

Let us discuss the second approximation made in the derivation of
eq.(8) which enables us to perform calculations in the
quasiclassical limit. In order to neglect the
{\it zitterbewegung} we must satisfy the condition
\cite{Te/JETP(90)}:
$${{\hbar}\over{2E_{\nu}}}|<{\dot {\vec O}}>|\ll 1. \eqno(21)$$
In application to the case of the Standard Model interactions it can
be rewritten as
$${{G_{F}}\over{E_{\nu}}}n_{eff}\ll 1. \eqno(22)$$
Here again the
background matter is taken to be nonmoving and unpolarized. This
condition means that quantum effects are not important in the
neutrino scattering off the background matter. Indeed, if, for
simplicity, we consider only electron component of the background
matter (the electron plasma) the constraint (22) is expressed in the
following way:
$$L^{2}\ll \lambda\sqrt{\sigma}, \eqno(23)$$
where $\sigma$
is the neutrino-electron cross section,
$\lambda\sim{{1}\over{n_{e}\sigma}}$ is the mean free
path of neutrino in this medium, $n_{e}$ is the electrons density. If
the condition (23) is satisfied, the characteristic dimensions,
$\lambda$ and $\sigma$,
corresponding to the scattering process
are much
larger than neutrino wave package width $L$. In the opposite case the
neutrino-electron scattering have to be treated within the
substantially quantum approach.

In summary we have derived the quasiclassical Lorentz invariant
equation for the neutrino spin evolution in the case when neutrinos
interact with external fields through a set of the most general
non-derivative interactions starting from the neutrino Lagrangian
that accounts for the scalar, pseudoscalar, vector, axial-vector,
tensor, pseudotensor terms. The advantage of the suggested new method
for derivation of the neutrino spin evolution equation is
manifested by the applicability to any theoretical
model which predicts the mentioned above neutrino interactions.
Furthermore the approach used enables us to examine the constraints
for the characteristics of matter and neutrino under which the
quasiclassical approach to the neutrino spin evolution is valid. As
an example we have obtained the neutrino spin evolution equation
for the Standard Model neutrino interactions in the case of moving
and polarized matter.

\end{document}